\documentclass[prd,amssymb,amsmath,showpacs,floatfix]{revtex4}
\usepackage{graphicx}
\usepackage{amssymb}
\usepackage{amsmath}

\newcommand{\be}{\begin{equation}}
\newcommand{\ee}{\end{equation}}
\newcommand{\bea}{\begin{eqnarray}}
\newcommand{\eea}{\end{eqnarray}}

\begin{document}

\title{ Measure of the Path Integral in Lattice Gauge Theory }

\author{F.~Paradis$^{a}$, H.~Kr\"{o}ger$^{a}$$\footnote{Corresponding author, Email: hkroger@phy.ulaval.ca}$, X.Q.~ Luo$^{b}$, K.J.M.~Moriarty$^{b}$  }

\affiliation{
$^{a}$ {\small\sl D\'{e}partement de Physique, Universit\'{e} Laval, Qu\'{e}bec, Qu\'{e}bec G1K 7P4, Canada} \\
$^{b}$ {\small\sl Department of Physics, Zhongshan University, 
Guangzhou 510275, People's Republic of China} \\
$^{c}$ {\small\sl Department of Mathematics, Statistics
and Computer Science, Dalhousie University, Halifax N.S. B3H 3J5, Canada}
}

\begin{abstract}
We show how to construct the measure of the path integral in lattice gauge theory. This measure contains a factor beyond the standard Haar measure. Such factor becomes relevant for the calculation of a single transition amplitude (in contrast to the calculation of ratios of amplitudes). Single amplitudes are required for computation of the partition function and the free energy. For ${\it U(1)}$ lattice gauge theory, we present a numerical simulation of the transition amplitude comparing the path integral with the evolution in terms of the Hamiltonian, showing good agreement.  
\end{abstract}


\maketitle




\section{Introduction}
\label{sec:Intro} 
In lattice gauge theory it is customary to compute the expectation value of an observable $\hat{O}[U]$, such as the Wilson loop, which is given by a ratio of Euclidean path integrals   
\begin{equation}
\label{eq:ExpObserv}
<\Omega,t=+\infty|\hat{O}[U]|\Omega,t=-\infty > = 
\frac{  \int [dU] ~ \hat{O}[U] ] ~ \exp[ -S[U]/\hbar ] }
{ \int [dU] ~ \exp[ -S[U]]/\hbar ] } ~ .
\end{equation}
Here $[dU]$ denotes the Haar measure of the group of gauge symmetry, like for example, $U(1)$, $SU(2)$, $SU(3)$, and $U$ denotes the link variables, being elements of such group. The state $\Omega$ denotes the vacuum. The reason why it is customary to compute such ratio of path integrals is due to the Monte Carlo method with importance sampling (like Metropolis) which works only for such ratio. In contrast, let us consider a single transition amplitude, like
\begin{equation}
\label{eq:Amplitude}
<U_{fi},t=T|U_{in},t=0 > = 
\int [dU] ~ \exp[ -S[U]/\hbar ]|_{U_{in},0}^{U_{fi},T} ~ .
\end{equation}
The states $|U_{in}>$, $|U_{fi}>$ denote a Bargman link states, which are defined by assigning a value $U_{ij}$ to each link $ij$ on the lattice in a fixed time slice. Now the measure $[dU]$ is no longer given by the Haar measure only, but there is a factor $Z^{N}$ involved for a lattice of 
$N-1$ intermediate time slices. It is the objective of this article to construct such measure. Physical scenarios which require the calculation of single amplitudes are: amplitudes of decay reactions, the partition function $Z(\beta)$, the free energy $F(\beta)$ given in terms of the partition function, or matrix elements involved in the construction of the Monte Carlo Hamiltonian \cite{Jirari99,Huang02}.

\section{Measure in 1-D quantum mechanics} 
Consider a Lagrangian of the form
\begin{equation}
L(x,\dot{x}) = \frac{1}{2} m \dot{x}^{2} - V(x) ~ .
\end{equation}
The transition amplitude in imaginary time, expressed as a 
path integral over $N-1$ intermediate time slices is given by
%
\begin{eqnarray}
&& \langle x_{fi}| \exp[-HT/\hbar] | x_{in} \rangle
= \lim_{N \to \infty}
\int_{-\infty}^{+\infty} dx_{1} \cdots dx_{N-1} 
\left( \sqrt{\frac{m}{2 \pi \hbar a_{0}} } \right)^{N}
\nonumber \\
&\times& \exp \left[ -\frac{1}{\hbar} \sum_{j=0}^{N-1} 
a_{0}[\frac{m}{2}(\frac{x_{j+1} - x_{j}}{a_{0}})^{2} 
- V(x_{j}) ]  \right] , x_{in} = x_{0}, ~ x_{fi} = x_{N}, ~ a_{0} = T/N ~ .  
\end{eqnarray}
%
The measure
\begin{equation} 
\label{eq:QMMeasure}
d \mu = dx_{1} \cdots dx_{N-1} 
\left( \sqrt{\frac{m}{2 \pi \hbar a_{0}} } \right)^{N} 
\end{equation}
contains the physical parameters of mass $m$, Planck's constant $\hbar$ and the length of a time slice $a_{0}$, but is independent of the potential. It can be obtained by computing the propagator of the kinetic term for a single time slice 
\begin{equation}
\langle x_{fi}| \exp[-\frac{\hat{P}^{2} a_{0}}{2m\hbar} ] | x_{in} \rangle 
= \left( \sqrt{\frac{m}{2 \pi \hbar a_{0}} } \right)
\exp \left[ -\frac{m}{2a_{0}\hbar}  
(x_{fi} - x_{in})^{2}   \right]
~ .
\end{equation}
This propagator has the following properties.
First, when $a_{0}$ goes to zero, the propagator goes over to
$\delta(x_{fi} - x_{in})$. The r.h.s. actually is a representation of the $\delta$ function by a Gaussian.  
Second, for any value of $a_{0}$ and $x_{in}$ one has
\begin{equation}
\label{eq:QMPropProbInterpret}
\int_{-\infty}^{+\infty} d x_{fi} 
\left( \sqrt{\frac{m}{2 \pi \hbar a_{0}} } \right)
\exp \left[ -\frac{m}{2a_{0}\hbar}  
(x_{fi} - x_{in})^{2}   \right]
= 1 ~ .
\end{equation}
Its interpretation is that the free propagator is the solution of a stochastic process (diffusion), and represents the probability for a random walker starting at $x_{in}$ at $t=0$ to arrive at $x_{fi}$ at time $t=a_{0}$. Summed over all possible final destinations, the probability must be one.

\section{Measure in lattice gauge theory}
Let us consider ${\it QED}$ on the lattice without fermions.
The group of gauge symmetry is ${\it U(1)}$. We keep in mind how to construct a lattice Hamiltonian (see e.g, Creutz\cite{Creutz}, Rothe\cite{Rothe}) via the transfer matrix, by splitting the lattice action into terms of time-like and space-like plaquettes, 
\begin{eqnarray}
\label{eq:SplitGaugeAction}
S[U] &=& \frac{1}{g^{2}}\frac{a}{a_{0}} \sum_{\Box_{time-like}} 
[1 -Re(U_{\Box})] + 
\frac{1}{g^{2}}\frac{a_{0}}{a} \sum_{\Box_{space-like}} 
[1 -Re(U_{\Box})] 
\nonumber \\
&\equiv&  S^{kin}[U] + S^{pot}[U] ~ .
\end{eqnarray}
Like in quantum mechanics, the extra factor in the measure 
is determined solely by the kinetic term. Thus we consider the single transition amplitude
\begin{equation}
\label{eq:FreeTransAmpl}
<U_{fi}|\exp[- H^{kin} T/\hbar ]| U_{in}> ~ , 
\end{equation}
where $H^{kin}$ denotes the kinetic term of the Kogut-Susskind lattice Hamiltonian \cite{Rothe}, given by
\begin{equation}
\label{eq:KogutSusskind}
H = \frac{g^{2}\hbar^{2}}{2a} \sum_{<ij>} \hat{l}_{ij}^{2} + \frac{1}{g^{2}a} \sum_{\Box_{space-like}} [1 - Re(U_{\Box})] 
\equiv H^{kin} + H^{pot} ~ .
\end{equation}
The Hamiltonian requires to choose a gauge and the Kogut-Susskind Hamiltonian has been obtained using the temporal gauge 
($U_{time-like}=1$). 
The propagator, Eq.(\ref{eq:FreeTransAmpl}), expressed as path integral reads
\begin{eqnarray}
&&\langle U_{fi}|\exp[- H^{kin} T/\hbar ]| U_{in} \rangle
\nonumber \\
&=& \left. \int Z [dU] ~ \exp \left[-\frac{1}{\hbar} S^{kin}[U] \right] \right|_{U_{in},t=0}^{U_{fi},t=T} 
\nonumber \\
&=& \left. \lim_{N \to \infty}
\int [\prod_{<ij>} Z_{ij}^{N} \prod_{k=1}^{N-1} dU_{ij}^{(k)} ]
~ \exp \left[- \frac{a}{\hbar g^{2} a_{0}} \sum_{k=0}^{N-1} \sum_{<ij>}
[1 -Re(U_{ij}^{(k)} (U_{ij}^{(k+1)})^{\dagger})] \right]
\right|_{U_{in},t=0}^{U_{fi},t=T} 
\nonumber \\
&=& \prod_{<ij>} \left( \left. \lim_{N \to \infty}
\int [ Z_{ij}^{N} \prod_{k=1}^{N-1} dU_{ij}^{(k)} ]
~ \exp \left[- \frac{a}{\hbar g^{2} a_{0}} \sum_{k=0}^{N-1} 
[1 -Re(U_{ij}^{(k)} (U_{ij}^{(k+1)})^{\dagger})] \right]
\right|_{U^{in}_{ij},t=0}^{U^{fi}_{ij},t=T} \right) ~ .
\nonumber \\
\end{eqnarray}
The amplitude factorizes into independent amplitudes 
for each spatial link $ij$. In order to compute the factor $Z$, 
let us consider a single link and its time evolution for a single 
time step ($T=a_{0}$). 
\begin{equation}
\label{eq:AmplSLinkSTimeStep}
\langle U_{fi}|\exp[- H^{kin} a_{0}/\hbar ]| U_{in} \rangle
= Z ~ \exp \left[- \frac{a}{\hbar g^{2} a_{0}}  
[1 - \cos(\alpha_{fi} - \alpha_{in})] \right] ~ .
\end{equation}
Like in quantum mechanics also in lattice gauge theory 
the Euclidean propagator has a probabilistic interpretation, 
and the analogue of Eq.(\ref{eq:QMPropProbInterpret}) holds,
\begin{equation}
\label{eq:LGTPropProbInterpret}
\int dU_{fi} \langle U_{fi}|\exp[- H^{kin} a_{0}/\hbar ]| U_{in} \rangle 
= \int_{- \pi}^{+\pi} \frac{d \alpha_{fi} }{2\pi} 
Z ~ \exp \left[- \frac{a}{\hbar g^{2} a_{0}}  
[1 - \cos(\alpha_{fi} - \alpha_{in})] \right] 
= 1 ~ .
\end{equation}
Defining $A= a/(\hbar g^{2}a_{0})$, and using the Bessel function of imaginary argument \cite{Gradshteyn},
\begin{equation}
I_{0}(z) = \frac{1}{\pi} \int_{0}^{\pi} d\theta \exp[z\cos(\theta)] ~ ,
\end{equation}
Eq.(\ref{eq:LGTPropProbInterpret}) yields
\begin{equation}
Z(A) = \frac{\exp(A)}{I_{0}(A)} ~ .
\end{equation}
For fixed $T$, $\lim_{N\to\infty}$ means $a_{0}\to 0$ and $A\to \infty$. Then one finds
\begin{equation}
\label{eq:ZFactor}
Z(A) = \sqrt{2 \pi A} [1 -\frac{1}{8} A^{-1} - \frac{7}{128} A^{-2} 
+ O(A^{-3})] ~ .
\end{equation}
In the limit $a_{0}\to 0$ the leading term $Z(A) = \sqrt{2 \pi A}$ is sufficient to guarantee that the amplitude in 
Eq.(\ref{eq:AmplSLinkSTimeStep}) goes over to $\delta(U_{fi} - U_{in})$, as should be. In quantum mechanics the leading term 
$Z(A) = \sqrt{ A/(2\pi)}$ with $A=m/(\hbar a_{0})$ is the exact result (see Eq.(\ref{eq:QMMeasure})). However, in lattice gauge theory, the sub-leading terms are important and can not be neglected. This has been confirmed by a numerical simulation discussed below.

\section{Comparison of propagator from path integral and Hamiltonian time evolution - a numerical simulation}
Let us consider a lattice which in spatial direction has a single link $ij$
and in time direction has $N$ time slices. We compare the propagator 
expressed via Hamiltonian time evolution with the path integral, using the measure of the group integral taking into account the previously calculted factor $Z$. 
\begin{eqnarray}
\label{eq:PathIntMultiSlice}
&&\langle U_{fi}|\exp[- H^{kin} T/\hbar ]| U_{in} \rangle 
\nonumber \\
&=& \left. Z^{N} \int_{- \pi}^{+\pi} [ \prod_{k=1}^{N-1} 
\frac{d \alpha^{(k)} }{2\pi} ]
~ \exp \left[- \frac{a}{\hbar g^{2} a_{0}} \sum_{k=0}^{N} 
[1 - \cos(\alpha^{(k)} - \alpha^{(k+1)})] \right] \right|_{\alpha^{(0)}=\alpha_{in}}^{\alpha^{(N)}=\alpha_{fi}} ~ .
\end{eqnarray}
\begin{figure*}[ht]
\begin{center}
\includegraphics[scale=0.60,angle=0]{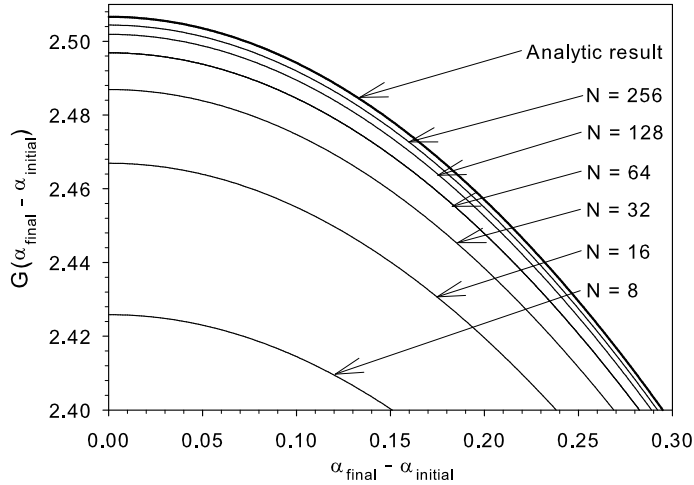}
\end{center}
\caption{Transition amplitude for single link state as function 
of initial and final link, $U_{in}=\exp(i\alpha_{in})$ and 
$U_{fi}=\exp(i\alpha_{fi})$. Transition time $T=1$, lattice spacing 
$a=1$. Comparison of Hamiltonian time 
evolution, Eq.(\ref{eq:HamTimeEvol}) (bold line) with path integral, 
Eq.(\ref{eq:PathIntMultiSlice}) (thin lines, number of time-slices $N=8,16,32,\dots,256$). 
 } 
\label{fig:TransAmpl}
\end{figure*}
\begin{figure*}[ht]
\begin{center}
\includegraphics[scale=0.60,angle=0]{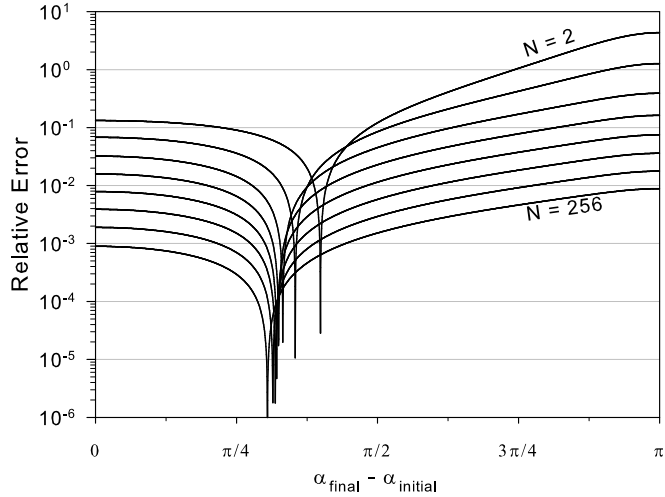}
\end{center}
\caption{Relative error of transition amplitude shown as function of $N=2,4,8,\dots,256$ and $\alpha_{fi}-\alpha_{in}$.}  
\label{fig:RelErrTransAmpl}
\end{figure*}
The propagator expressed by Hamiltonian time evolution is given by
\begin{equation}
\label{eq:HamTimeEvol}
<U_{fi}|\exp[- H^{kin} T/\hbar ]| U_{in}> 
= \sum_{n=0,\pm1,\pm2,\dots} 
\exp[- \frac{g^{2} \hbar T}{2a} n^{2} ]
\cos[n(\alpha_{in} - \alpha_{fi})] ~ .
\end{equation}
This has been calculated using the basis of eigen states of the electric field operator
\begin{equation}
\label{eq:ElectField}
\hat{l}_{ij} |\lambda_{ij}> = \lambda_{ij} |\lambda_{ij}> ~ , \lambda_{ij} = 0,\pm 1, \pm 2, \dots 
\end{equation}
and the connection from the Bargmann link basis to the electric field string basis given by the scalar product 
\begin{equation}
\label{eq:ScalarProd}
<\lambda|U> = (U)^{\lambda} ~ .
\end{equation}
The path integral, Eq.(\ref{eq:PathIntMultiSlice}), has been evaluated by numerical integration using 10000 mesh points in each time slice. The results as function of the angle $\alpha_{fi}-\alpha_{in}$
and of the number of time slices $N$ is shown in 
Fig.[\ref{fig:TransAmpl}]. One observes convergence when increasing 
$N=8,16,32,\dots,256$. The relative error, shown in 
Fig.[\ref{fig:RelErrTransAmpl}] varies between $10^{-2}$ and $10^{-3}$ depending on the angle and decreases monotonically when increasing $N=2,4,8,\dots,256$. On the other hand, when taking into account only the leading term in Eq.(\ref{eq:ZFactor}) the relative error was found to be in the order of $10 - 20\%$.

\vspace*{6pt}

\noindent {\bf Acknowledgement.} H.Kr\"oger and K.J.M. Moriarty have been supported by NSERC Canada. X.Q. Luo has been supported by Key Project of NSF China (10235040), and by National and Guangdong Ministries of Education.

\end{document}